\documentclass[12pt]{article}
\usepackage{graphicx,color,epsfig,rotate}
\title{ Valence transition behavior of the doped Falicov-Kimball
model at nonzero temperatures}
\author{Pavol Farka\v sovsk\'y \\
Institute  of  Experimental  Physics,  Slovak   Academy   of
Sciences\\
Watsonova 47, 043 53 Ko\v {s}ice, Slovakia}
\date{}
\begin{document}
\baselineskip=20pt
\maketitle

\begin{abstract}
The extrapolation of small-cluster exact-diagonalization calculations
is used to study the influence of doping on valence transitions in the 
spinless Falicov-Kimball model at nonzero temperatures. Two types of 
doping are examined, and namely, the substitution of rare-earth ions by 
non-magnetic ions that introduce (i) one or (ii) none additional electron 
(per non-magnetic ion) into the conduction band. It is found that the first 
type of substitution increases the average $f$-state occupancy of rare-earth 
ions, whereas the second type of substitution has the opposite effect. 
The results obtained are used to describe valence transition behavior 
of samarium  in the hexaboride solid solutions
$Sm_{1-x}M_xB_6$ ($M=Y^{3+},La^{3+}, Sr^{2+},Yb^{2+}$)
and a very good agreement of theoretical and experimental results 
is found.
\end{abstract}
\newpage

In the past decade, a considerable amount of effort has  been devoted 
to  understanding  of  the  multitude  of  anomalous physical 
properties of rare-earth and transition-metal compounds.
Recent theoretical works based on exact numerical and analytical calculations 
showed that many of these anomalous features,  e.g., mixed  valence phenomena, 
metal-insulator transitions, etc., can be described very well within different 
versions of the Falicov-Kimball model~(FKM)~\cite{Falicov,Cho,Free}.
In particular, it has been found that the spinless FKM, can describe the 
discontinuous valence and insulator-metal transitions~\cite{Fark1,Fark2} 
(induced by external pressure) observed experimentally in some rare-earth 
sulfides, halides and borides~\cite{Wachter}. A nice correspondence between 
theoretical and experimental results has been found, for example, 
for $SmB_6$~\cite{Fark2,Cool,Park,Fark3}, where the alternative explanation of 
formation and disappearance of the energy gap in the electronic spectrum 
of $SmB_6$ has been presented. 
For $SmB_6$ a very good agreement of experimental and theoretical results
has been obtained also for nonzero temperatures for both the temperature
dependence of the specific heat as well as magnetic 
susceptibility~\cite{Macedo}. In addition, it was found that the spinless 
FKM can describe the anomalous temperature increase of electrical 
conductivity $\sigma$ in $SmB_6$ (due to the valence transition)~\cite{Fark4} 
as well as the low temperature behavior of $\sigma$ (due to the appearance 
of the in-gap structure) in this material~\cite{Fark3,Batko}. 
Recently, the FKM has been successfully used to describe the inelastic 
light scattering in the $SmB_6$ systems~\cite{Raman}.
These results indicate that the spinless FKM is a quite good microscopic 
model that can describe, at least qualitatively, many of anomalous 
features of rare-earth compounds. 
From this point of view it seems to be interesting to ask if this simple model 
can describe also another kind of experiments performed often in these 
materials, and namely, the influence of doping on the  valence transition 
behavior of systems. For  $SmB_6$ it has been found, for
example~\cite{Tarascon,Gabani}, that the substitution of $Sm$ 
by non-magnetic divalent ions (e.g., $Sr^{2+},Yb^{2+}$) increases the average 
samarium valence (the average occupancy of $f$ orbitals decreases), whereas 
the substitution of $Sm$ by non-magnetic trivalent ions (e.g., $Y^{3+},La^{3+}$) 
produces the opposite effect. In this paper we try to describe the influence 
of both types of substitutions on the average samarium valence within 
the spinless FKM.  

The spinless FKM is based on the coexistence of two different
types of electronic states in given materials: localized, highly correlated
ioniclike states and extended, uncorrelated, Bloch-like states. It is
generally accepted that valence and insulator-metal transitions result 
from a change in the occupation numbers of these electronic states, 
which remain themselves basically unchanged in their character.
The Hamiltonian of the model can be written as the sum of three
terms:
\begin{equation}
H=\sum_{ij}t_{ij}d^+_id_j+U\sum_if^+_if_id^+_id_i+E_f\sum_if^+_if_i,
\end{equation}
where $f^+_i$, $f_i$ are the creation and annihilation
operators  for an electron in the localized $f$ state 
of a rare-earth atom at lattice site $i$ with binding energy 
$E_f$ and $d^+_i$,
$d_i$ are the creation and annihilation operators
of the itinerant spinless electrons in the $d$-band
Wannier state at site $i$.
The first term of (1) is the kinetic energy corresponding to
quantum-mechanical hopping of the itinerant $d$ electrons
between sites $i$ and $j$. These intersite hopping
transitions are described by the matrix  elements $t_{ij}$,
which are $-t$ if $i$ and $j$ are the nearest neighbors and
zero otherwise (in the following all parameters are measured
in units of $t$). The second term represents the on-site
Coulomb interaction between the $d$-band electrons 
(with the total number $N_d=\sum_id^+_id_i$) and the localized 
$f$ electrons (with the total number $N_f=\sum_if^+_if_i$). 
The third  term stands for the localized $f$ electrons whose 
sharp energy level is $E_f$.
The Hamiltonian of the FKM doped by divalent (e.g., $Sr^{2+}$)
or trivalent (e.g., $Y^{3+}$) ions has the same form. From the numerical
point of view there is only one difference, and namely, the summation
over all lattice sites (in the second and third term) should be replaced 
by summation over the samarium positions.

Since in the spinless version of the FKM
without hybridization  the $f$-electron occupation
number $f^+_if_i$ of each site $i$ commutes with
the Hamiltonian (1), the $f$-electron occupation number
is a good quantum number, taking only two values: $w_i=1$
or 0, according to whether or not the site $i$ is occupied
by the localized $f$ electron. Then the Hamiltonian (1) can be written as

\begin{equation}
H=\sum_{ij}h_{ij}d^+_id_j+E_f\sum_iw_i,
\end{equation}
where $h_{ij}(w)=t_{ij}+Uw_i\delta_{ij}$.

Thus for a given $f$-electron configuration
$w=\{w_1,w_2 \dots w_L\}$ defined on the one or two-di\-men\-sional
lattice of $L$ sites with periodic boundary conditions, the Hamiltonian (2)
is the second-quantized version of the single-particle
Hamiltonian $h(w)=T+UW$,  where $T$ is an $L$-square matrix
with elements $t_{ij}$ and $W$ is the diagonal matrix with
elements $w_i$. 
The behavior of this model at nonzero temperatures is 
studied using small-cluster exact-diagonalization
calculations. In order to compensate partially for the small
size of clusters, thermal properties of the system are
investigated via the grand canonical ensemble. 

The grand canonical partition function of the FKM~\cite{Kennedy}
can be expressed directly in terms of the eigenvalues $\epsilon_i$ of
the single-particle operator $h$, which depend on the $f$-electron
configuration specified by $w=\{w_i\}$, i.e.,

\begin{equation}
\Xi=\sum_{w}e^{-(E_f-\mu)N_f/\tau}\prod_{i=1}^L[1+
e^{-(\epsilon_i-\mu)/\tau}].
\end{equation}

The temperature dependences ($\tau=k_BT$) of the thermodynamic quantities 
are calculated from this expression by using
the standard statistical mechanics procedure. For example,
the total number of $f$-electrons  $N_f$ and the internal energy
$(\cal E)$ as functions of chemical potential $\mu$ are
given by~\cite{Brandt_Schmidt}

\begin{eqnarray}
N_f(\mu)&=&-{\tau}\frac{\partial}{\partial E_f}\ln \Xi  ,\\
\cal E(\mu)&=&\frac{1}{L}(\mu\tau\frac{\partial}{\partial\mu}-\frac{\partial}
{\partial\beta})\ln \Xi,
\end{eqnarray}
where $\beta=1/\tau$.

In the next, we consider (i) the substitution of rare-earth ions 
(e.g., $Sm$) by non-magnetic trivalent ions (e.g., $Y^{3+}$) which 
introduce conduction electrons into the $d$-conduction band (one electron 
per dopant) and (ii) the substitution of rare-earth ions by non-magnetic
divalent ions (e.g., $Sr^{2+}$) which play a dilution role and reduce 
the number of conduction electrons in the $d$-conduction band 
(no additional electrons are introduced to the system). 
In the first case the total number of electrons $N=N_f+N_d$ is equal to 
the total number of lattice sites $L$ (the case of one electron 
per rare-earth atom is considered) while in the second case $N=L-X$,
where $X$ is the total number of dopants.
Thus the first step of numerical calculations is to determine
the chemical potential corresponding to $N=L$ and $N=L-X$, respectively. 
Afterwards $n_f, \cal E$  and  other thermodynamic quantities can be 
calculated numerically.

It should be noted that the summation over all $f$ electron distributions 
in (3) should be replaced for the doped FKM by summation over all possible  
$f$-electron and dopant distributions, thereby strongly limiting
numerical computations. Although the number of configurations
can be reduced considerably by the use of symmetries of $H$,
there is still a limit $(L \sim 20)$ on the size of clusters
that can be studied with this method. However, we will show
later that due to small sensitivity of the thermodynamic
quantities of the FKM on $L$, already such small clusters can 
describe the thermodynamics of the model very well.

To describe effects of doping by divalent and trivalent ions on the 
average $f$-state occupancy of rare-earth ions we have performed 
an exhaustive study of the model for a wide range of model parameters 
$U,\tau, E_f$. The typical dependences of the average $f$-orbital 
occupancy $n_f=N_f/(L-X)$ on the dopant concentration $x=X/L$ are displayed 
in Fig.~1 for $E_f=0, U=0.5$ and several different temperatures $\tau$.

\begin{figure}[htb]
\hspace{1cm}
\includegraphics[angle=0,width=12cm,scale=1]{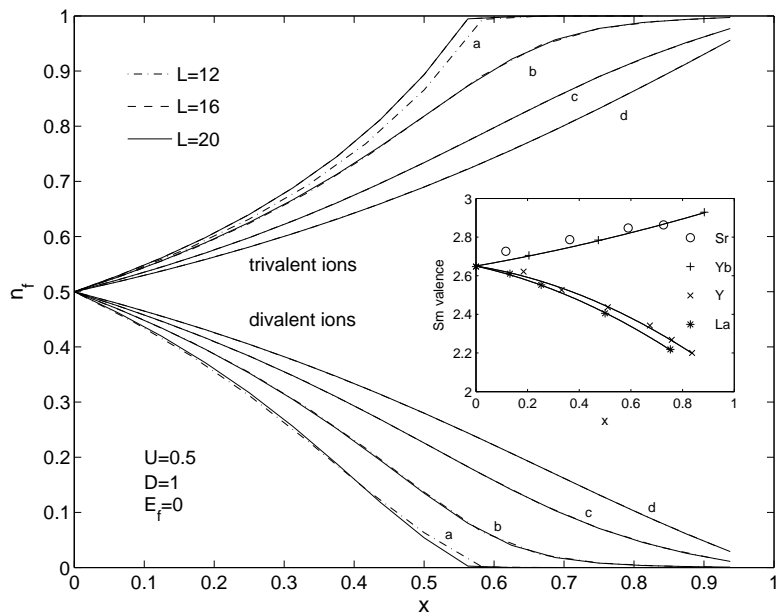}
\caption{
The average occupancy of $f$ orbitals as a function of $x$
for doping by divalent and trivalent ions calculated for 
$L=12,16,20$ and several different temperatures $\tau$.
Curve $a$, $\tau=0.1$; curve $b$, $\tau=0.3$;  
curve $c$, $\tau=0.6$; curve $d$, $\tau=1$.  
Inset: The average samarium valence in the $Sm_{1-x}M_xB_6$ systems 
$(M=Y^{3+},La^{3+}, Sr^{2+},Yb^{2+})$ measured at 300 K (Ref.~14).}
\label{fig1}
\end{figure}

To determine the finite-size effects all concentration
dependences have been calculated for several cluster sizes
($L=12,16,20$) and displayed on the common linear scale.
We have found that for all examined temperatures $\tau$
the finite-size effects are negligible (over the whole range
of $x$ plotted) and thus these results can be used satisfactorily 
to represent the behavior of macroscopic systems.
It is seen that the substitution of rare-earth ions by trivalent
and divalent ions produces fully different effects.
While in the first case the average occupancy of $f$-orbitals increases
with increasing dopant concentration $x$, in the second case $n_f$  
decreases. In all examined cases the valence transition realizes
continuously. Comparing these results with experimental measurements 
of the average samarium valence $v$ (see inset 
in Fig.~1) in the  $Sm_{1-x}M_xB_6$ systems~\cite{Tarascon,Gabani} 
($M=Y^{3+},La^{3+}, Sr^{2+},Yb^{2+}$) one can find a good qualitative 
agreement between the theoretical and experimental results. 
Indeed,  the substitution of $Sm$ by non-magnetic divalent ions 
($Sr^{2+},Yb^{2+}$) increases the average samarium valence 
(the average occupancy of $f$ orbitals $n_f=3-v$ decreases),
whereas the substitution of $Sm$ by non-magnetic trivalent ions 
($Y^{3+},La^{3+}$) produces the opposite effect. 
\begin{figure}[tb]
\hspace{1cm}
\includegraphics[angle=0,width=12cm,scale=1]{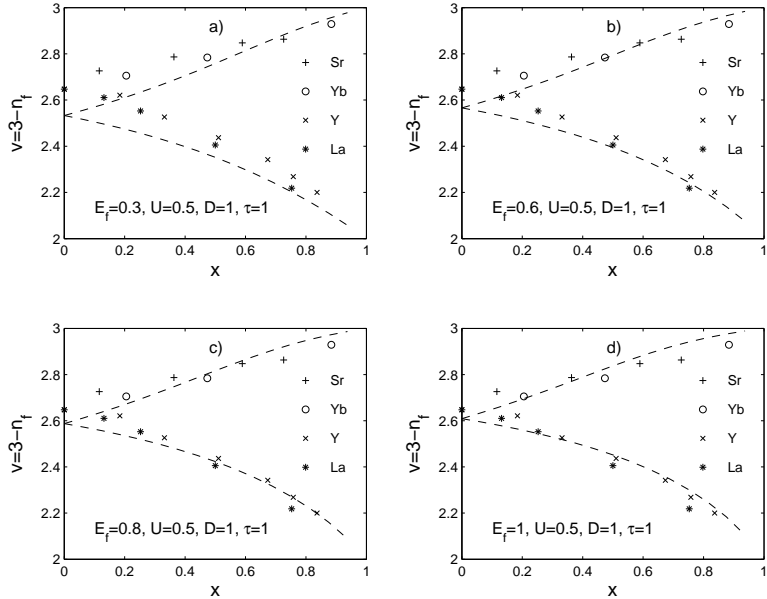}
\caption{
Comparison of the one-dimensional theoretical (dashed lines) and experimental 
results for the average $Sm$ valence $v=3-n_f$ as a function of $x$.
The theoretical results have been calculated for $U=0.5, \tau=1$ 
and several different values of the $f$-level position $E_f$.
The finite-size effects are negligible, on the linear scale it is not
possible on the drawing to distinguish behaviors obtained for $L=12, 16$ 
and 20 over the whole range of $x$ plotted.
The experimental results have been obtained for T=300 K.}
\label{fig2}
\end{figure}
These results again indicate that the FKM, in spite of its relative 
simplicity could, in principle, yield the correct physics for describing 
rare-earth compounds. Even, also small quantitative deviations between the 
experimental and theoretical results can be straightforwardly improved
within the FKM by optimizing the $n_f(x)$ behavior with respect to
$\tau$ and $E_f$. This is illustrated in Fig.~2, where the 
$x$-dependence of the $f$-orbital occupancy is plotted for $\tau=1$
(this relatively large value is chosen to ensure that $\tau$ is 
much greater than the spacing of many-body energy-levels for both $D=1$
and $D=2$) and several different values of $E_f$. 
\begin{figure}[tb]
\hspace{1cm}
\includegraphics[angle=0,width=12cm,scale=1]{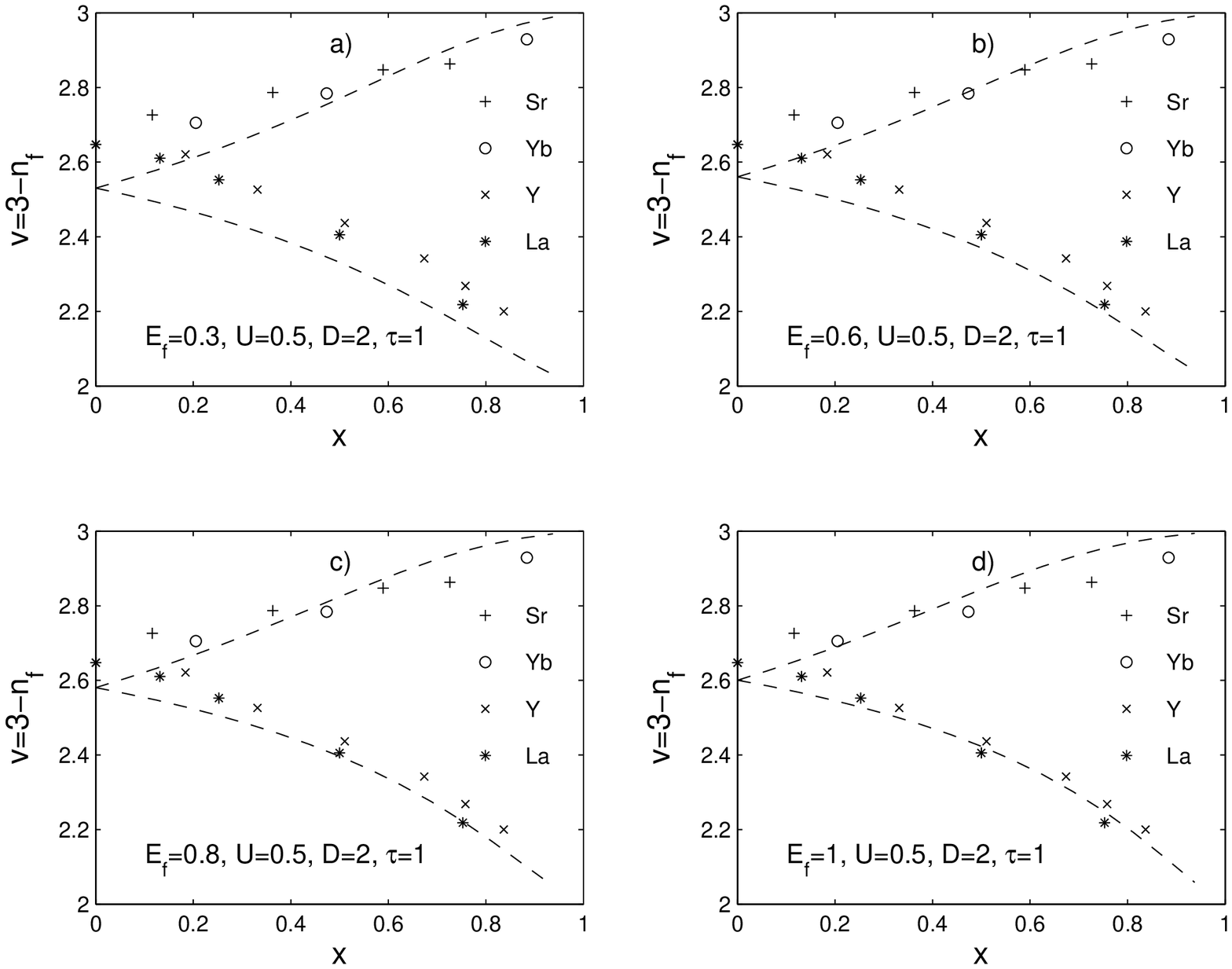}
\caption{
Comparison of the two-dimensional theoretical (dashed lines) and 
experimental results for the average $Sm$ valence $v=3-n_f$ as a function 
of $x$. The theoretical results have been obtained on clusters with 
$L=10,16,20$  for the same set of $U, \tau$ and $E_f$ values as in $D=1$. 
Again, the finite-size effects are negligible.} 
\label{fig3}
\end{figure}
It is seen that for nonzero 
values of $E_f$ a nice correspondence between the theoretical and 
experimental results can be achieved. Of course, one can take objection   
that real materials are three dimensional, while our results have 
been obtained for the one dimensional case. Unfortunately, due to
limitations (mentioned above) on the size of clusters that can be 
treated exactly by small-cluster exact-diagonalization calculations
($L=20$) we are not able to obtain any exact results in three dimensions.
However, in two dimensions there are several finite clusters with
$L=10,16,18,20$ that are accessible to exact-diagonalization
calculations~\cite{Dag} and can be used to test the stability of obtained 
results with increasing dimension. 
The numerical calculations that we have performed in two dimensions 
(for the same set of model parameters as in D=1 and the above mentioned 
cluster sizes) showed (see Fig.~3) that the $f$-occupancy
of rare-earth orbitals depends only very weakly on the dimension of the
system. Thus we can conclude that the spinless FKM, when treated exactly,
can provide a reasonable theoretical description of valence transitions
induced by doping in some rare-earth compounds, e.g.,
the samarium hexaboride solid solutions
$Sm_{1-x}M_xB_6$ ($M=Y^{3+},La^{3+}, Sr^{2+},Yb^{2+}$).
Of course, this statement can not be extended automatically on all
thermodynamics characteristics of the doped FKM. For example, to do 
quantitative comparison between the theoretical and experimental results
for the specific heat and the magnetic susceptibility, the model should 
be first generalized by including the spin and orbital dynamics that 
description was too simplified in the Hamiltonian~(1). The work in this
direction is currently in progress.
 
In summary, the extrapolation of small-cluster exact-diagonalization 
calculations was used to study the influence of doping on valence 
transitions in the spinless FKM at nonzero temperatures. 
Two types of doping were examined, and namely, the substitution of 
rare-earth ions by non-magnetic ions that introduce (i) one or (ii) none 
additional electron (per non-magnetic ion) into the conduction band. 
It was found that the first type of substitution increases the average 
$f$-state occupancy of rare-earth ions, whereas the second type of 
substitution has the opposite effect. In all examined cases valence
changes are continuous. The results obtained were used to describe valence 
transition behavior of samarium  in the hexaboride solid solutions
$Sm_{1-x}M_xB_6$ ($M=Y^{3+},La^{3+}, Sr^{2+},Yb^{2+}$)
and a very good agreement of theoretical and experimental results 
was found.

\vspace{0.35cm}

{\bf Acknowledgments}

\vspace{0.35cm}
This work was supported by the Slovak Grant Agency VEGA under Grant
No. 2/4060/04 and the Science and Technology Assistance Agency under 
Grant APVT-20-021602.

\end{document}